\documentclass[aps,prl,twocolumn,superscriptaddress]{revtex4-1}

\usepackage[dvips]{epsfig}

\usepackage[sort&compress]{natbib}

\begin{document}


\title{Simultaneous wavelength translation and amplitude modulation of single photons from a quantum dot}


\author{Matthew T. Rakher} \email{matthew.rakher@gmail.com}
\altaffiliation{Current Address: Departement Physik, Universit\"{a}t
Basel, Klingelbergstrasse 82, CH-4056 Basel, Switzerland}
\affiliation{Center for Nanoscale Science and Technology, National
Institute of Standards and Technology, Gaithersburg, MD 20899, USA}
\author{Lijun Ma}
\affiliation{Information Technology Laboratory, National
Institute of Standards and Technology, Gaithersburg, MD 20899, USA}
\author{Marcelo Davan\c{c}o}
\affiliation{Center for Nanoscale Science and Technology, National
Institute of Standards and Technology, Gaithersburg, MD 20899, USA}
\affiliation{Maryland NanoCenter, University of Maryland, College Park, MD 20742, USA}
\author{Oliver Slattery}
\affiliation{Information Technology Laboratory, National
Institute of Standards and Technology, Gaithersburg, MD 20899, USA}
\author{Xiao Tang}
\affiliation{Information Technology Laboratory, National
Institute of Standards and Technology, Gaithersburg, MD 20899, USA}
\author{Kartik Srinivasan}
\affiliation{Center for Nanoscale Science and Technology, National
Institute of Standards and Technology, Gaithersburg, MD 20899, USA}


\date{\today}

\begin{abstract}
Hybrid quantum information devices that combine disparate physical
systems interacting through photons offer the promise of combining
low-loss telecommunications wavelength transmission with high
fidelity visible wavelength storage and manipulation. The
realization of such systems requires control over the waveform of
single photons to achieve spectral and temporal matching. Here, we
experimentally demonstrate the simultaneous wavelength conversion
and temporal shaping of single photons generated by a quantum dot
emitting near 1300 nm with an exponentially-decaying waveform
(lifetime $\approx$1.5~ns). Quasi-phase-matched sum-frequency
generation with a pulsed 1550~nm laser creates single photons at
710~nm with a controlled amplitude modulation at 350 ps timescales.

\end{abstract}

\pacs{}

\maketitle

The integration of disparate quantum systems is an ongoing effort in
the development of distributed quantum
networks~\cite{ref:Kimble_Nat08}.  Two challenges in hybrid schemes
which use photons for coupling include the differences in transition
frequencies and linewidths among the component systems. Previous
experiments using non-linear optical media to translate (or
transduce) photons from one wavelength to another while preserving
quantum-mechanical
properties~\cite{ref:Huang_Kumar_PRL,*ref:Tanzili_Zbinden,*ref:McGuinnes_PRL10,ref:Rakher_NPhot_2010}
provide a means to overcome the first impediment.  The second
constraint has been addressed through single photon waveform
manipulation using direct electro-optic
modulation~\cite{ref:Kolchin_PRL_08,*ref:Specht_NPhot_09,*ref:Rakher_APL_2011},
$\Lambda$-type cavity-QED~\cite{ref:McKeever,*ref:Keller} and atomic
ensemble systems~\cite{ref:Eisaman_PRL04,*ref:Chen_PRL10}, and
conditional, non-local operations in spontaneous parametric
downversion~\cite{ref:Baek_PRA08}.  Here, we use pulsed frequency
upconversion to simultaneously change the frequency and temporal
amplitude profile of single photons produced by a semiconductor
quantum dot (QD).  Triggered single photons that have an
exponentially decaying temporal profile with a time constant of 1.5
ns and a wavelength of 1300 nm are converted to photons that have a
Gaussian temporal profile with a controllable full-width at
half-maximum (FWHM) as narrow as 350 ps $\pm$~16~ps and a wavelength
of 710 nm. The use of a high conversion efficiency nonlinear waveguide and low-loss fiber optics results in a 16~$\%$ overall efficiency in producing such frequency converted, amplitude-modulated photons.  The simultaneous combination of wavepacket manipulation and quantum
frequency conversion may be valuable in integrating
telecommunications-band semiconductor QDs with broadband visible
wavelength quantum
memories~\cite{ref:Reim_NatPhot10,*ref:Saglamyurek_Nat11} as part of
a distributed quantum network, for the creation of ultra-high bandwidth~\cite{ref:Kielpinski_PRL_2011}, indistinguishable photons from disparate quantum emitters, and for efficient, temporally-resolved photon counting at the ps level.

\begin{figure}[t]
\centering
\includegraphics[width=8cm, clip=true]{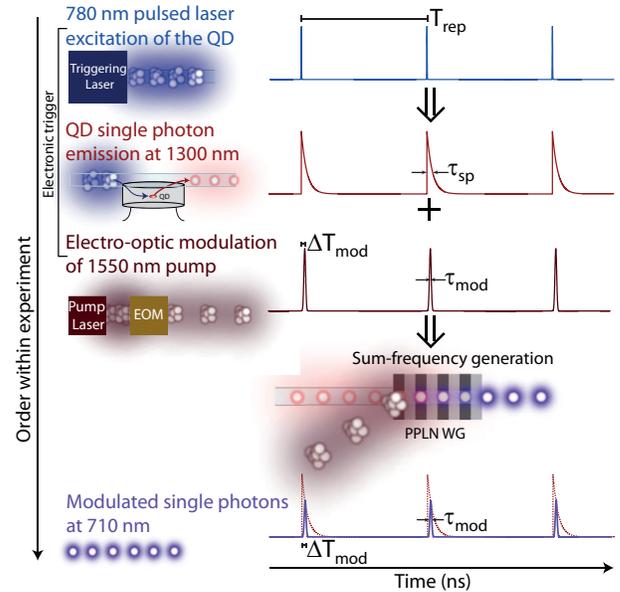}
\caption{Schematic of the experiment for simultaneous wavelength translation and amplitude modulation of single photons from a quantum dot.}
    \label{fig:fig1}
\end{figure}

Single epitaxially-grown semiconductor QDs are promising stable,
bright, and scalable on-demand sources of single
photons~\cite{ref:Shields_NPhot}, with improvements in photon
extraction efficiency~\cite{ref:Strauf_NPhot,ref:Claudon},
suppression of multi-photon
events~\cite{ref:Claudon,ref:Ates_PRL09}, and photon
indistinguishability~\cite{ref:Santori2,ref:Ates_PRL09} indicative
of their potential for high performance in quantum information
applications.  On the other hand, the dominant and most mature system for such QD single photon source developments has been the InGaAs/GaAs material system, whose band structure constrains the range of available emission energies, and temporal control of the photon
wavepacket shape remains a challenge in these systems.  Despite
recent progress~\cite{ref:Fernandez_PRL}, access to three-level
Raman transitions in which the temporal shape is determined by the
pump mode profile, a staple of trapped atom and ion
systems~\cite{ref:McKeever,ref:Keller,ref:Vasilev} is typically not
available.  Instead, most QDs are two-level systems in which the
emitted photons have an exponentially decaying temporal profile, and
temporal shaping must occur after photon generation.  As we describe
below, the method used in this work produces wavelength-translated, single photon
wavepackets with a temporal profile inherited from the classical
pump beam used in the frequency upconversion process. Though
this technique is not lossless, it can still be quite efficient, is flexible, straightforward to
use on existing devices, and operates on the nanosecond timescales
requisite for QD experiments and for which classical coherent pulse
shaping techniques~\cite{ref:Weiner_RSI} are difficult to implement.  In comparison to direct amplitude modulation of single photon wavepackets~\cite{ref:Kolchin_PRL_08,*ref:Specht_NPhot_09,*ref:Rakher_APL_2011}, the technique presented here has lower loss, can operate on much faster timescales using existing mode-locked laser technologies, and importantly, simultaneously changes the wavelength of the photons.  This is necessary for integration with visible wavelength quantum systems and provides a method to overcome spectral and temporal distinguishability of disparate quantum sources.

\begin{figure}[h]
\centering
\includegraphics[width=8cm, clip=true]{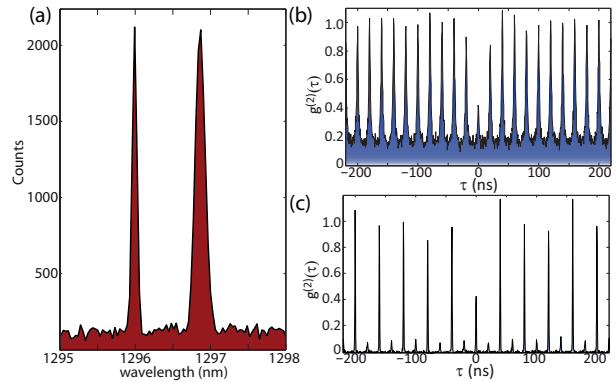}
\caption{(a) PL spectrum of a single QD after 60~s integration
showing two excitonic transitions, $X^+$ (1296~nm) and $X^0$ (1297
nm).  (b-c) Second-order intensity correlation, $g^{(2)}(\tau)$, of
single photons upconverted with a CW pump, where
$g^{2}(0)=0.41\pm0.02$ (c) and a pulsed pump with
$\tau_{mod}$=500~ps, where $g^{2}(0)=0.45\pm0.03$ (d) for an
integration time of 7200~s.} \label{fig:fig2}
\end{figure}

We generated single photons at 1.3~$\mu$m from a single InAs QD
embedded in a GaAs mesa~\cite{SM_PRL1}.  The QD sample is cooled to a
temperature of $\approx$7 K, excited with a repetitively pulsed
(50 MHz) laser, and its photoluminescence (PL) is out-coupled into a single mode fiber as depicted in Fig.~\ref{fig:fig1}. The
PL is directed either into a grating spectrometer for characterization or into the
pulsed upconversion setup for simultaneous wavelength translation and amplitude modulation.  The PL spectrum from a single QD measured by the
spectrometer is shown in Fig.~\ref{fig:fig2}(a). It displays two sharp peaks corresponding to two
excitonic charge configurations, $X^+$ near 1296 nm, and $X^0$ near
1297 nm.  Photons emitted at the $X^0$ transition wavelength will be
used for the experiments described hereafter.

PL from the $X^0$ transition is directed into an upconversion setup
where it is combined with a strong 1550~nm pulse in a
periodically-poled LiNbO$_3$ (PPLN) waveguide~\cite{SM_PRL1}.
A simplified schematic of the experimental timing sequence is shown
in Fig.~\ref{fig:fig1}. The pump pulse is created by gating the
output of a tunable laser with an electro-optic modulator (EOM).  An
electrical pulse generator drives the EOM synchronously with the
780~nm QD excitation laser, but at half the repetition rate (25
MHz), using a trigger signal from the delay generator.  These
instruments combine to generate electrical pulses with controllable
FWHM ($\tau_{mod}$) and delay ($\Delta T_{mod}$) as shown in
Fig.~\ref{fig:fig1}, and the resulting optical pulses have an
extinction ratio $>20$~dB.  The modulated 1550~nm pump signal is amplified to produce a peak power of 85~mW entering the PPLN waveguide where it interacts with a 1300 nm QD single photon to create a
single photon at the sum-frequency near 710~nm.  This $\chi^{(2)}$
process is made efficient through quasi-phase-matching by periodic
poling \cite{ref:Fejer_IEEE} as well as the tight optical
confinement of the waveguide \cite{ref:Chanvillard}. Previous
measurements using a continuous-wave (CW) pump in the same setup
demonstrated single-photon conversion efficiencies $\gtrsim~75$~$\%$ \cite{ref:Rakher_NPhot_2010}, and others have
measured efficiencies near unity with attenuated laser pulses
\cite{ref:Langrock_Fejer,ref:Vandevender_Kwiat_JMO,*ref:Albota_Wong_upconversion,*ref:Xu_Tang}.
Light exiting the PPLN is spectrally filtered to isolate the 710 nm
photons, which are detected by Si single photon counting avalanche
detectors (SPADs) for excited state lifetime and second-order
correlation measurement ($g^{(2)}(\tau)$).

First, we compare the measured $g^{(2)}(\tau)$ for photons that are
upconverted using a 1550~nm CW pump (Fig.~\ref{fig:fig2}(b)) and
500~ps pulses (Fig.~\ref{fig:fig2}(c)) with the same peak power of 85 mW. Both are antibunched with
$g^{(2)}(0)<0.5$, showing that the signal is dominantly composed of
single photons in both cases.  However, pulsed pumping reduces
events that are uncorrelated in time with the QD single photons and
contribute a constant background. This unwanted background results
from upconversion of anti-Stokes Raman photons from the strong (CW)
1550 nm beam~\cite{ref:Langrock_Fejer}, and is seen in
Fig.~\ref{fig:fig2}(b) but not in Fig.~\ref{fig:fig2}(c). For
understanding the origin of the non-zero $g^{(2)}(0)$ value, the
background is helpful in distinguishing the fraction due to
anti-Stokes Raman photons from that due to upconversion of
multi-photon emission from the QD
sample~\cite{ref:Rakher_NPhot_2010}. For a practical implementation,
however, it adds a constant level to the communications channel and pulsed upconversion removes this
noise without gating of the detector. Ideally,
Fig.~\ref{fig:fig2}(c) would only show peaks spaced by 40 ns, due to
the 25 MHz repetition rate of the EOM. In practice, the small peaks
spaced 20 ns from the large peaks are due to imperfect extinction of
the EOM and pulse generator, resulting in upconversion of QD
photons when the EOM is nominally off.

\begin{figure}[h]
\centering
\includegraphics[width=8cm, clip=true]{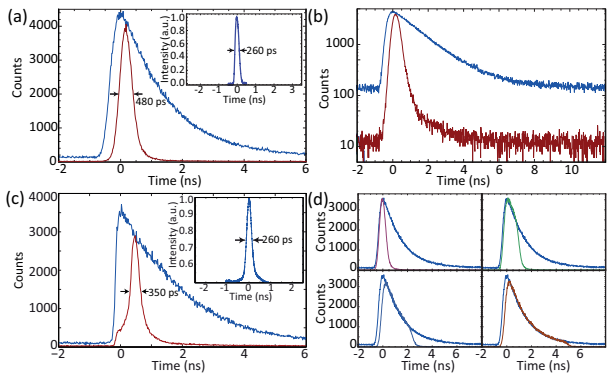}
\caption{(a,b) Temporal profile of the upconverted photons using a
CW (blue) and $\tau_{mod}=$260~ps pulsed (maroon) 1550 nm pump on
linear (a) and log (b) scales. Inset: 1550~nm pump pulse
measured by the optical communications analyzer.  (c) Same as (a)
but using a reduced timing jitter SPAD. (d)
Temporal profile of upconverted photons using $\tau_{mod}=\{0.5, 1.25, 2.5, 5.1\}$~ns along with a CW pump.  All measurements are taken with 1200~s integration.}
\label{fig:fig3}
\end{figure}

Next, we perform time-resolved measurements of the upconverted
710~nm photons. In recent work~\cite{ref:Rakher_NPhot_2010} using a
CW 1550 nm pump beam, the temporal amplitude profile of the
upconverted 710~nm photon exactly matched that of the original
1300~nm photon, and was used to measure the QD lifetime with
dramatically better dynamic range than with a telecommunications SPAD. Here,
the pulsed 1550~nm pump not only upconverts the QD photon to 710~nm,
but also modulates its temporal amplitude profile because
$\tau_{mod}$ is less than the lifetime of the QD transition
($\approx1.5$~ns). Figure~\ref{fig:fig3}(a) displays the temporal
amplitude profile of 710~nm single photons generated using a 1550~nm
pulse with $\tau_{mod}=260$~ps (maroon), along with that of single
photons generated with a CW pump (blue) for comparison.  The
measured 480~ps~$\pm$~16~ps FWHM of the upconverted photon is
limited by the $\approx$350 ps timing jitter of the Si SPAD and its
uncertainty is given by the timebase of the photon counting board.
The same plot is reproduced in Fig.~\ref{fig:fig3}(b) on a log
scale, with an apparent increase in the dynamic range due to the removal of CW anti-Stokes Raman photons.  This same measurement was
performed using a SPAD with a reduced timing jitter
($\approx50$~ps), and the resulting data is shown in
Fig.~\ref{fig:fig3}(c) corresponding to a FWHM of
350~ps~$\pm$~16~ps.  Here, the resulting FWHM is not limited by the detector timing jitter but by an effective broadening of the pump pulse in the frequency conversion process~\cite{SM_PRL1}. Even so,
taken together with the commercial availability of 40 GHz EOMs and
drivers for 1550 nm lasers, these results demonstrate a first step
towards the creation of quantum light sources that are modulated to
operate near the capacity of telecommunications
channels~\cite{ref:Kielpinski_PRL_2011}.  To show the versatility of
the setup, Fig.~\ref{fig:fig3}(d) shows the temporal profile of QD
single photons after upconversion using pump pulse widths of
$\tau_{mod}=$ 500~ps, 1.25~ns, 2.5~ns, and 5.1~ns
along with a CW pump for comparison.  By simply adjusting
the pulse generator that drives the EOM, one can create single
photons of arbitrary width and shape~\cite{SM_PRL1}.

\begin{figure}[h]
\centering
\includegraphics[width=8cm, clip=true]{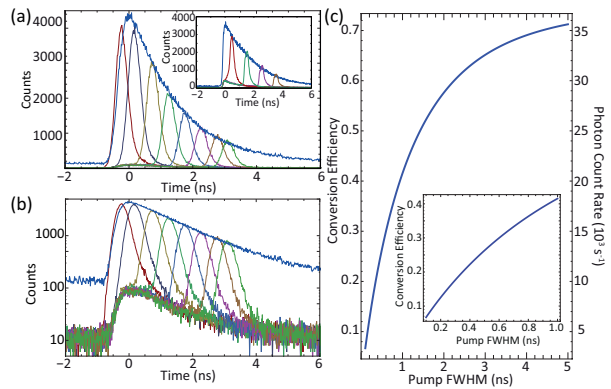}
\caption{(a,b) Temporal profile of the upconverted photons using a
CW (blue) and $\tau_{mod}=$260~ps pulsed (various colors) 1550 nm
pump on linear (a) and log (b) scales for delays $\Delta
T_{mod}=\{0.0, 0.5, 1.0, 1.5, 2.0, 2.5, 3.0, 3.35\}$~ns.  Inset:
Similar measurement as (a), but measured with reduced timing jitter
SPAD.  All measurements are taken with
1200~s integration. (c) Net conversion efficiency and photon count rate as a function of the pump pulse FWHM, $\tau_{mod}$.  Inset: Focus on sub-ns regime.} \label{fig:fig4}
\end{figure}

In addition to changing $\tau_{mod}$, the delay between the arrival
of the QD single photon and pump pulse, $\Delta T_{mod}$, can also
be varied (Fig.~\ref{fig:fig1}). Figure~\ref{fig:fig4}(a)-(b)
show the result of such a measurement in linear (a) and log (b)
scale for $\Delta T_{mod}=\{0.0, 0.5, 1.0, 1.5, 2.0, 2.5, 3.0,
3.35\}$~ns under pulsed pumping with $\tau_{mod}=260$~ps.  The inset
of (a) shows a similar measurement using the reduced timing jitter
SPAD.  The peaks heights nicely follow the decay curve of the CW
profile, shown in blue for comparison. This measurement suggests that pulsed frequency
upconversion could be used for achieving high timing resolution in
experiments on single quantum emitters.  These time-correlated
single photon counting experiments are currently limited by the
timing jitter of the SPAD, which is typically $>50$~ps. The
time-domain sampling enabled by pulsed
upconversion~\cite{ref:Shah_JQE,*ref:Ma_OE_11} provides a timing
resolution set by $\tau_{mod}$, which is limited by the quasi-phase-matching spectral bandwidth of the non-linear material. For the PPLN
waveguide used here, the bandwidth ($\approx0.35$~nm) corresponds to
a minimum $\tau_{mod}\approx$10~ps, while sub-ps timing resolution
should be available in broader bandwidth
systems~\cite{ref:kuzucu_wong,*ref:Suchowski_OPN_10}. Sub-ps 1550 nm pulses
can be generated by mode-locked fiber lasers, and if applied as in
Fig.~\ref{fig:fig4}(a)-(b), could trace out emission dynamics with a
timing resolution 1-2 orders of magnitude better than the typical
timing jitter of a SPAD, allowing, for example, measurement of beat
phenomena within the QD~\cite{ref:Flissikowski} or time-domain observation of vacuum Rabi oscillations in cavity quantum electrodynamics~\cite{ref:Srinivasan17}.

The data from Figs.~\ref{fig:fig2} and
\ref{fig:fig3} indicate that while the quantum
nature of the photon has been inherited from the QD emission near
1300~nm, its temporal profile has been inherited from the strong
pump pulse near 1550~nm. This is a direct consequence of the nonlinear nature of the upconversion process. However, because QD-generated single photons have a coherence
time that is typically less than twice the lifetime, they are not
perfectly indistinguishable~\cite{ref:Santori2}.  This arises due to interaction of the confined carriers in the QD with
the surrounding crystal and, for the type of QD considered here,
yields a coherence time of
$\approx280$~ps~\cite{ref:Srinivasan17} and an indistinguishability
of $\approx10$~$\%$.  For our experiments, this means that each
photon is not modulated in the same way and the resulting
histograms are ensemble averages. Nonetheless, the experiments would
proceed in the exact same manner for more ideal photons, such as those
produced with a ns coherence time through near-resonant excitation~\cite{ref:Ates_PRL09}.  In fact, simultaneous frequency translation and amplitude modulation can be used to generate indistinguishable single photons from non-identical QDs \cite{ref:patel_Nphot2010,*ref:Flagg_PRL10}.  Frequency translation can move each single photon to the same wavelength while amplitude modulation can be used to select the coherent part of the wave-packet.  Since the quasi-phase-matching of the PPLN can be tuned by temperature, this offers the ability to create indistinguishable single photons from QDs spectrally separated by the entire inhomogeneous linewidth of the QD distribution (which is usually tens of nanometers) without the need for electrical gates or modification of the sample.

The single photon manipulation demonstrated here is essentially a
combination of quantum frequency conversion and amplitude
modulation.  Coherent pulse-shaping
techniques~\cite{ref:Weiner_RSI}, which have been used with
entangled photon pairs~\cite{ref:Silberberg_PRL}, are currently
quite difficult to directly apply to QD single photons due to their
narrow spectral bandwidth compared to photons produced by parametric
downconversion, for example. Furthermore, recent
work~\cite{ref:Kielpinski_PRL_2011} has suggested that a combination
of frequency upconversion using a spectrally tailored 1550 nm pump
beam and spectral phase correction may be an approach to lossless
shaping of QD single photons. Our work, utilizing a similar sum
frequency generation approach, represents a step towards such a
goal.  Though our approach is not lossless, broadband insertion loss
(usually $>3$ dB) is avoided in comparison to direct amplitude
modulation of the single photon state because the modulation in
nonlinear mixing approaches such as ours and that of
Ref.~\cite{ref:Kielpinski_PRL_2011} is performed on the classical
pump beam.  Nonetheless, the fact that the pump pulse is temporally shorter than the single photon wave-packet introduces extra loss.  A full derivation of this loss is included in the supplemental material~\cite{SM_PRL1}, but the result is shown in Fig.~\ref{fig:fig4}(c) which plots the net conversion efficiency as a function of the pump pulse FWHM, $\tau_{mod}$, from 100~ps to 5~ns (inset displays sub-ns regime).  The efficiency asymptotically approaches 75~$\%$, the measured conversion efficiency with a CW pump, and ranges from 16~$\%$ for $\tau_{mod}=260$~ps to 71~$\%$ for $\tau_{mod}=5$~ns.  For our FTW-based PL collection with 0.1~$\%$ collection efficiency and 50 MHz excitation rate, this translates to a single photon count rate of 8000~$s^{-1}$ to 36000 $s^{-1}$ as shown on the right axis in Fig.~\ref{fig:fig4}(c).  Using more advanced techniques that have demonstrated $>$10~$\%$ collection efficiency \cite{ref:Strauf_NPhot,*ref:Claudon}, the overall production rate of frequency translated, temporally modulated single photons can easily reach $10^6$~$s^{-1}$.

In summary, we have demonstrated simultaneous wavelength translation and amplitude modulation of a single photon from a quantum dot using pulsed frequency upconversion.  The use of a quasi-phase-matched waveguide and low-loss fiber optics results in a 16~$\%$ overall conversion efficiency in producing Gaussian-shaped, single photon pulses near 710~nm with a FWHM of 350~ps from single photons near 1.3~$\mu$m with an exponentially-decaying wavepacket.  Such methods may prove valuable for integrating disparate quantum systems, creating ultra-high bandwidth indistinguishable
single photon sources, and for achieving high resolution in
time-resolved experiments of single quantum systems.

%


\begin{thebibliography}{40}%
\makeatletter
\providecommand \@ifxundefined [1]{%
 \@ifx{#1\undefined}
}%
\providecommand \@ifnum [1]{%
 \ifnum #1\expandafter \@firstoftwo
 \else \expandafter \@secondoftwo
 \fi
}%
\providecommand \@ifx [1]{%
 \ifx #1\expandafter \@firstoftwo
 \else \expandafter \@secondoftwo
 \fi
}%
\providecommand \natexlab [1]{#1}%
\providecommand \enquote  [1]{``#1''}%
\providecommand \bibnamefont  [1]{#1}%
\providecommand \bibfnamefont [1]{#1}%
\providecommand \citenamefont [1]{#1}%
\providecommand \href@noop [0]{\@secondoftwo}%
\providecommand \href [0]{\begingroup \@sanitize@url \@href}%
\providecommand \@href[1]{\@@startlink{#1}\@@href}%
\providecommand \@@href[1]{\endgroup#1\@@endlink}%
\providecommand \@sanitize@url [0]{\catcode `\\12\catcode `\$12\catcode
  `\&12\catcode `\#12\catcode `\^12\catcode `\_12\catcode `\%12\relax}%
\providecommand \@@startlink[1]{}%
\providecommand \@@endlink[0]{}%
\providecommand \url  [0]{\begingroup\@sanitize@url \@url }%
\providecommand \@url [1]{\endgroup\@href {#1}{\urlprefix }}%
\providecommand \urlprefix  [0]{URL }%
\providecommand \Eprint [0]{\href }%
\providecommand \doibase [0]{http://dx.doi.org/}%
\providecommand \selectlanguage [0]{\@gobble}%
\providecommand \bibinfo  [0]{\@secondoftwo}%
\providecommand \bibfield  [0]{\@secondoftwo}%
\providecommand \translation [1]{[#1]}%
\providecommand \BibitemOpen [0]{}%
\providecommand \bibitemStop [0]{}%
\providecommand \bibitemNoStop [0]{.\EOS\space}%
\providecommand \EOS [0]{\spacefactor3000\relax}%
\providecommand \BibitemShut  [1]{\csname bibitem#1\endcsname}%
\let\auto@bib@innerbib\@empty
\bibitem [{\citenamefont {{Kimble}}(2008)}]{ref:Kimble_Nat08}%
  \BibitemOpen
  \bibfield  {author} {\bibinfo {author} {\bibfnamefont {H.~J.}\ \bibnamefont
  {{Kimble}}},\ }\href {\doibase 10.1038/nature07127} {\bibfield  {journal}
  {\bibinfo  {journal} {Nature (London)}\ }\textbf {\bibinfo {volume} {453}},\
  \bibinfo {pages} {1023} (\bibinfo {year} {2008})}\BibitemShut {NoStop}%
\bibitem [{\citenamefont {Huang}\ and\ \citenamefont
  {Kumar}(1992)}]{ref:Huang_Kumar_PRL}%
  \BibitemOpen
  \bibfield  {author} {\bibinfo {author} {\bibfnamefont {J.~M.}\ \bibnamefont
  {Huang}}\ and\ \bibinfo {author} {\bibfnamefont {P.}~\bibnamefont {Kumar}},\
  }\href@noop {} {\bibfield  {journal} {\bibinfo  {journal} {Phys. Rev. Lett.}\
  }\textbf {\bibinfo {volume} {68}},\ \bibinfo {pages} {2153} (\bibinfo {year}
  {1992})}\BibitemShut {NoStop}%
\bibitem [{\citenamefont {Tanzilli}\ \emph {et~al.}(2005)\citenamefont
  {Tanzilli}, \citenamefont {Tittel}, \citenamefont {Halder}, \citenamefont
  {Alibart}, \citenamefont {Baldi}, \citenamefont {Gisin},\ and\ \citenamefont
  {Zbinden}}]{ref:Tanzili_Zbinden}%
  \BibitemOpen
  \bibfield  {author} {\bibinfo {author} {\bibfnamefont {S.}~\bibnamefont
  {Tanzilli \emph {et~al.}}},\ }\href@noop {} {\bibfield  {journal} {\bibinfo
   {journal} {Nature}\ }\textbf {\bibinfo {volume} {437}},\ \bibinfo {pages}
  {116} (\bibinfo {year} {2005})}\BibitemShut {NoStop}%
\bibitem [{\citenamefont {McGuinness}\ \emph {et~al.}(2010)\citenamefont
  {McGuinness}, \citenamefont {Raymer}, \citenamefont {McKinstrie},\ and\
  \citenamefont {Radic}}]{ref:McGuinnes_PRL10}%
  \BibitemOpen
  \bibfield  {author} {\bibinfo {author} {\bibfnamefont {H.~J.}\ \bibnamefont
  {McGuinness \emph {et~al.}}},\ }\href {\doibase 10.1103/PhysRevLett.105.093604} {\bibfield
  {journal} {\bibinfo  {journal} {Phys. Rev. Lett.}\ }\textbf {\bibinfo
  {volume} {105}},\ \bibinfo {pages} {093604} (\bibinfo {year}
  {2010})}\BibitemShut {NoStop}%
\bibitem [{\citenamefont {Rakher}\ \emph {et~al.}(2010)\citenamefont {Rakher},
  \citenamefont {Ma}, \citenamefont {Slattery}, \citenamefont {Tang},\ and\
  \citenamefont {Srinivasan}}]{ref:Rakher_NPhot_2010}%
  \BibitemOpen
  \bibfield  {author} {\bibinfo {author} {\bibfnamefont {M.~T.}\ \bibnamefont
  {Rakher \emph {et~al.}}},\ }\href {\doibase
  10.1038/nphoton.2010.221} {\bibfield  {journal} {\bibinfo  {journal} {Nature
  Photonics}\ }\textbf {\bibinfo {volume} {4}},\ \bibinfo {pages} {786}
  (\bibinfo {year} {2010})}\BibitemShut {NoStop}%
\bibitem [{\citenamefont {Kolchin}\ \emph {et~al.}(2008)\citenamefont
  {Kolchin}, \citenamefont {Belthangady}, \citenamefont {Du}, \citenamefont
  {Yin},\ and\ \citenamefont {Harris}}]{ref:Kolchin_PRL_08}%
  \BibitemOpen
  \bibfield  {author} {\bibinfo {author} {\bibfnamefont {P.}~\bibnamefont
  {Kolchin \emph {et~al.}}},\ }\href
  {\doibase 10.1103/PhysRevLett.101.103601} {\bibfield  {journal} {\bibinfo
  {journal} {Phys. Rev. Lett.}\ }\textbf {\bibinfo {volume} {101}},\ \bibinfo
  {pages} {103601} (\bibinfo {year} {2008})}\BibitemShut {NoStop}%
\bibitem [{\citenamefont {Specht}\ \emph {et~al.}(2009)\citenamefont {Specht},
  \citenamefont {Bochmann}, \citenamefont {M{\"u}cke}, \citenamefont {Weber},
  \citenamefont {Figueroa}, \citenamefont {Moehring},\ and\ \citenamefont
  {Rempe}}]{ref:Specht_NPhot_09}%
  \BibitemOpen
  \bibfield  {author} {\bibinfo {author} {\bibfnamefont {H.~P.}\ \bibnamefont
  {Specht \emph {et~al.}}},\ }\href {\doibase
  10.1038/nphoton.2009.115} {\bibfield  {journal} {\bibinfo  {journal} {Nature
  Photonics}\ }\textbf {\bibinfo {volume} {3}},\ \bibinfo {pages} {469}
  (\bibinfo {year} {2009})}\BibitemShut {NoStop}%
\bibitem [{\citenamefont {{Rakher}}\ and\ \citenamefont
  {{Srinivasan}}(2011)}]{ref:Rakher_APL_2011}%
  \BibitemOpen
  \bibfield  {author} {\bibinfo {author} {\bibfnamefont {M.~T.}\ \bibnamefont
  {{Rakher}}}\ and\ \bibinfo {author} {\bibfnamefont {K.}~\bibnamefont
  {{Srinivasan}}},\ }\href {\doibase 10.1063/1.3593007} {\bibfield  {journal}
  {\bibinfo  {journal} {Appl. Phys. Lett.}\ }\textbf {\bibinfo {volume} {98}},\
  \bibinfo {pages} {211103} (\bibinfo {year} {2011})}\BibitemShut {NoStop}%
\bibitem [{\citenamefont {McKeever}\ \emph {et~al.}(2004)\citenamefont
  {McKeever}, \citenamefont {Boca}, \citenamefont {Boozer}, \citenamefont
  {Miller}, \citenamefont {Buck}, \citenamefont {Kuzmich},\ and\ \citenamefont
  {Kimble}}]{ref:McKeever}%
  \BibitemOpen
  \bibfield  {author} {\bibinfo {author} {\bibfnamefont {J.}~\bibnamefont
  {McKeever \emph {et~al.}}},\ }\href@noop {} {\bibfield
  {journal} {\bibinfo  {journal} {Science}\ }\textbf {\bibinfo {volume}
  {303}},\ \bibinfo {pages} {1992} (\bibinfo {year} {2004})}\BibitemShut
  {NoStop}%
\bibitem [{\citenamefont {Keller}\ \emph {et~al.}(2004)\citenamefont {Keller},
  \citenamefont {Lange}, \citenamefont {Hayaska}, \citenamefont {Lange},\ and\
  \citenamefont {Walther}}]{ref:Keller}%
  \BibitemOpen
  \bibfield  {author} {\bibinfo {author} {\bibfnamefont {M.}~\bibnamefont
  {Keller \emph {et~al.}}},\ }\href@noop {} {\bibfield
  {journal} {\bibinfo  {journal} {Nature}\ }\textbf {\bibinfo {volume} {431}},\
  \bibinfo {pages} {1075} (\bibinfo {year} {2004})}\BibitemShut {NoStop}%
\bibitem [{\citenamefont {Eisaman}\ \emph {et~al.}(2004)\citenamefont
  {Eisaman}, \citenamefont {Childress}, \citenamefont {Andr\'e}, \citenamefont
  {Massou}, \citenamefont {Zibrov},\ and\ \citenamefont
  {Lukin}}]{ref:Eisaman_PRL04}%
  \BibitemOpen
  \bibfield  {author} {\bibinfo {author} {\bibfnamefont {M.~D.}\ \bibnamefont
  {Eisaman \emph {et~al.}}},\ }\href {\doibase
  10.1103/PhysRevLett.93.233602} {\bibfield  {journal} {\bibinfo  {journal}
  {Phys. Rev. Lett.}\ }\textbf {\bibinfo {volume} {93}},\ \bibinfo {pages}
  {233602} (\bibinfo {year} {2004})}\BibitemShut {NoStop}%
\bibitem [{\citenamefont {Chen}\ \emph {et~al.}(2010)\citenamefont {Chen},
  \citenamefont {Zhang}, \citenamefont {Yan}, \citenamefont {Loy},
  \citenamefont {Wong},\ and\ \citenamefont {Du}}]{ref:Chen_PRL10}%
  \BibitemOpen
  \bibfield  {author} {\bibinfo {author} {\bibfnamefont {J.~F.}\ \bibnamefont
  {Chen \emph {et~al.}}},}\href {\doibase
  10.1103/PhysRevLett.104.183604} {\bibfield  {journal} {\bibinfo  {journal}
  {Phys. Rev. Lett.}\ }\textbf {\bibinfo {volume} {104}},\ \bibinfo {pages}
  {183604} (\bibinfo {year} {2010})}\BibitemShut {NoStop}%
\bibitem [{\citenamefont {Baek}\ \emph {et~al.}(2008)\citenamefont {Baek},
  \citenamefont {Kwon},\ and\ \citenamefont {Kim}}]{ref:Baek_PRA08}%
  \BibitemOpen
  \bibfield  {author} {\bibinfo {author} {\bibfnamefont {S.-Y.}\ \bibnamefont
  {Baek}}, \bibinfo {author} {\bibfnamefont {O.}~\bibnamefont {Kwon}}, \ and\
  \bibinfo {author} {\bibfnamefont {Y.-H.}\ \bibnamefont {Kim}},\ }\href
  {\doibase 10.1103/PhysRevA.77.013829} {\bibfield  {journal} {\bibinfo
  {journal} {Phys. Rev. A}\ }\textbf {\bibinfo {volume} {77}},\ \bibinfo
  {pages} {013829} (\bibinfo {year} {2008})}\BibitemShut {NoStop}%
\bibitem [{\citenamefont {{Reim}}\ \emph {et~al.}(2010)\citenamefont {{Reim}},
  \citenamefont {{Nunn}}, \citenamefont {{Lorenz}}, \citenamefont {{Sussman}},
  \citenamefont {{Lee}}, \citenamefont {{Langford}}, \citenamefont {{Jaksch}},\
  and\ \citenamefont {{Walmsley}}}]{ref:Reim_NatPhot10}%
  \BibitemOpen
  \bibfield  {author} {\bibinfo {author} {\bibfnamefont {K.~F.}\ \bibnamefont
  {{Reim \emph {et~al.}}}},}\href {\doibase
  10.1038/nphoton.2010.30} {\bibfield  {journal} {\bibinfo  {journal} {Nature
  Photonics}\ }\textbf {\bibinfo {volume} {4}},\ \bibinfo {pages} {218}
  (\bibinfo {year} {2010})}\BibitemShut {NoStop}%
\bibitem [{\citenamefont {{Saglamyurek}}\ \emph {et~al.}(2011)\citenamefont
  {{Saglamyurek}}, \citenamefont {{Sinclair}}, \citenamefont {{Jin}},
  \citenamefont {{Slater}}, \citenamefont {{Oblak}}, \citenamefont
  {{Bussi{\`e}res}}, \citenamefont {{George}}, \citenamefont {{Ricken}},
  \citenamefont {{Sohler}},\ and\ \citenamefont
  {{Tittel}}}]{ref:Saglamyurek_Nat11}%
  \BibitemOpen
  \bibfield  {author} {\bibinfo {author} {\bibfnamefont {E.}~\bibnamefont
  {{Saglamyurek \emph {et~al.}}}},}\href {\doibase
  10.1038/nature09719} {\bibfield  {journal} {\bibinfo  {journal} {Nature
  (London)}\ }\textbf {\bibinfo {volume} {469}},\ \bibinfo {pages} {512}
  (\bibinfo {year} {2011})}\BibitemShut {NoStop}%
\bibitem [{\citenamefont {Kielpinski}\ \emph {et~al.}(2011)\citenamefont
  {Kielpinski}, \citenamefont {Corney},\ and\ \citenamefont
  {Wiseman}}]{ref:Kielpinski_PRL_2011}%
  \BibitemOpen
  \bibfield  {author} {\bibinfo {author} {\bibfnamefont {D.}~\bibnamefont
  {Kielpinski}}, \bibinfo {author} {\bibfnamefont {J.~F.}\ \bibnamefont
  {Corney}}, \ and\ \bibinfo {author} {\bibfnamefont {H.~M.}\ \bibnamefont
  {Wiseman}},\ }\href {\doibase 10.1103/PhysRevLett.106.130501} {\bibfield
  {journal} {\bibinfo  {journal} {Phys. Rev. Lett.}\ }\textbf {\bibinfo
  {volume} {106}},\ \bibinfo {pages} {130501} (\bibinfo {year}
  {2011})}\BibitemShut {NoStop}%
\bibitem [{\citenamefont {{Shields}}(2007)}]{ref:Shields_NPhot}%
  \BibitemOpen
  \bibfield  {author} {\bibinfo {author} {\bibfnamefont {A.~J.}\ \bibnamefont
  {{Shields}}},\ }\href {\doibase 10.1038/nphoton.2007.46} {\bibfield
  {journal} {\bibinfo  {journal} {Nature Photonics}\ }\textbf {\bibinfo
  {volume} {1}},\ \bibinfo {pages} {215} (\bibinfo {year} {2007})}\BibitemShut
  {NoStop}%
\bibitem [{\citenamefont {{Strauf}}\ \emph {et~al.}(2007)\citenamefont
  {{Strauf}}, \citenamefont {{Stoltz}}, \citenamefont {{Rakher}}, \citenamefont
  {{Coldren}}, \citenamefont {{Petroff}},\ and\ \citenamefont
  {{Bouwmeester}}}]{ref:Strauf_NPhot}%
  \BibitemOpen
  \bibfield  {author} {\bibinfo {author} {\bibfnamefont {S.}~\bibnamefont
  {{Strauf \emph {et~al.}}}},\ }\href {\doibase 10.1038/nphoton.2007.227} {\bibfield
  {journal} {\bibinfo  {journal} {Nature Photonics}\ }\textbf {\bibinfo
  {volume} {1}},\ \bibinfo {pages} {704} (\bibinfo {year} {2007})}\BibitemShut
  {NoStop}%
\bibitem [{\citenamefont {{Claudon}}\ \emph {et~al.}(2010)\citenamefont
  {{Claudon}}, \citenamefont {{Bleuse}}, \citenamefont {{Malik}}, \citenamefont
  {{Bazin}}, \citenamefont {{Jaffrennou}}, \citenamefont {{Gregersen}},
  \citenamefont {{Sauvan}}, \citenamefont {{Lalanne}},\ and\ \citenamefont
  {{G{\'e}rard}}}]{ref:Claudon}%
  \BibitemOpen
  \bibfield  {author} {\bibinfo {author} {\bibfnamefont {J.}~\bibnamefont
  {{Claudon \emph {et~al.}}}},\ }\href {\doibase
  10.1038/nphoton.2009.287} {\bibfield  {journal} {\bibinfo  {journal} {Nature
  Photonics}\ }\textbf {\bibinfo {volume} {4}},\ \bibinfo {pages} {174}
  (\bibinfo {year} {2010})}\BibitemShut {NoStop}%
\bibitem [{\citenamefont {Ates}\ \emph {et~al.}(2009)\citenamefont {Ates},
  \citenamefont {Ulrich}, \citenamefont {Reitzenstein}, \citenamefont
  {L\"offler}, \citenamefont {Forchel},\ and\ \citenamefont
  {Michler}}]{ref:Ates_PRL09}%
  \BibitemOpen
  \bibfield  {author} {\bibinfo {author} {\bibfnamefont {S.}~\bibnamefont
  {Ates \emph {et~al.}}},\ }\href {\doibase
  10.1103/PhysRevLett.103.167402} {\bibfield  {journal} {\bibinfo  {journal}
  {Phys. Rev. Lett.}\ }\textbf {\bibinfo {volume} {103}},\ \bibinfo {pages}
  {167402} (\bibinfo {year} {2009})}\BibitemShut {NoStop}%
\bibitem [{\citenamefont {Santori}\ \emph {et~al.}(2002)\citenamefont
  {Santori}, \citenamefont {Fattal}, \citenamefont {Vuckovic}, \citenamefont
  {Solomon},\ and\ \citenamefont {Yamamoto}}]{ref:Santori2}%
  \BibitemOpen
  \bibfield  {author} {\bibinfo {author} {\bibfnamefont {C.}~\bibnamefont
  {Santori \emph {et~al.}}},\ }\href@noop {}
  {\bibfield  {journal} {\bibinfo  {journal} {Nature}\ }\textbf {\bibinfo
  {volume} {419}},\ \bibinfo {pages} {594} (\bibinfo {year}
  {2002})}\BibitemShut {NoStop}%
\bibitem [{\citenamefont {Fernandez}\ \emph {et~al.}(2009)\citenamefont
  {Fernandez}, \citenamefont {Volz}, \citenamefont {Desbuquois}, \citenamefont
  {Badolato},\ and\ \citenamefont {Imamoglu}}]{ref:Fernandez_PRL}%
  \BibitemOpen
  \bibfield  {author} {\bibinfo {author} {\bibfnamefont {G.}~\bibnamefont
  {Fernandez \emph {et~al.}}},\ }\href@noop {}
  {\bibfield  {journal} {\bibinfo  {journal} {Phys. Rev. Lett.}\ }\textbf
  {\bibinfo {volume} {103}},\ \bibinfo {pages} {087406} (\bibinfo {year}
  {2009})}\BibitemShut {NoStop}%
\bibitem [{\citenamefont {{Vasilev}}\ \emph {et~al.}(2010)\citenamefont
  {{Vasilev}}, \citenamefont {{Ljunggren}},\ and\ \citenamefont
  {{Kuhn}}}]{ref:Vasilev}%
  \BibitemOpen
  \bibfield  {author} {\bibinfo {author} {\bibfnamefont {G.~S.}\ \bibnamefont
  {{Vasilev}}}, \bibinfo {author} {\bibfnamefont {D.}~\bibnamefont
  {{Ljunggren}}}, \ and\ \bibinfo {author} {\bibfnamefont {A.}~\bibnamefont
  {{Kuhn}}},\ }\href@noop {} {\bibfield  {journal} {\bibinfo  {journal} {New
  Journal of Physics}\ }\textbf {\bibinfo {volume} {12}},\ \bibinfo {pages}
  {063024} (\bibinfo {year} {2010})}\BibitemShut {NoStop}%
\bibitem [{\citenamefont {{Weiner}}(2000)}]{ref:Weiner_RSI}%
  \BibitemOpen
  \bibfield  {author} {\bibinfo {author} {\bibfnamefont {A.~M.}\ \bibnamefont
  {{Weiner}}},\ }\href@noop {} {\bibfield  {journal} {\bibinfo  {journal}
  {Review of Scientific Instruments}\ }\textbf {\bibinfo {volume} {71}},\
  \bibinfo {pages} {1929} (\bibinfo {year} {2000})}\BibitemShut {NoStop}%
\bibitem [{SM_()}]{SM_PRL1}%
  \BibitemOpen
  \href@noop {} {}\bibinfo {note} {See Supplemental Material at [URL will be
  inserted by publisher] for experimental details, a full analysis of the total
  conversion efficiency, and more complex amplitude profiles.}\BibitemShut
  {Stop}%
\bibitem [{\citenamefont {{Fejer}}\ \emph {et~al.}(1992)\citenamefont
  {{Fejer}}, \citenamefont {{Magel}}, \citenamefont {{Jundt}},\ and\
  \citenamefont {{Byer}}}]{ref:Fejer_IEEE}%
  \BibitemOpen
  \bibfield  {author} {\bibinfo {author} {\bibfnamefont {M.~M.}\ \bibnamefont
  {{Fejer \emph {et~al.}}}},\ }\href {\doibase 10.1109/3.161322} {\bibfield  {journal}
  {\bibinfo  {journal} {IEEE J. Quan. Elec.}\ }\textbf {\bibinfo {volume}
  {28}},\ \bibinfo {pages} {2631} (\bibinfo {year} {1992})}\BibitemShut
  {NoStop}%
\bibitem [{\citenamefont {Chanvillard}\ \emph {et~al.}(2000)\citenamefont
  {Chanvillard}, \citenamefont {Aschi\'{e}ri}, \citenamefont {Baldi},
  \citenamefont {Ostrowsky}, \citenamefont {de~Micheli}, \citenamefont
  {Huang},\ and\ \citenamefont {Bamford}}]{ref:Chanvillard}%
  \BibitemOpen
  \bibfield  {author} {\bibinfo {author} {\bibfnamefont {L.}~\bibnamefont
  {Chanvillard \emph {et~al.}}},\ }\href {\doibase
  10.1063/1.125948} {\bibfield  {journal} {\bibinfo  {journal} {Appl. Phys.
  Lett.}\ }\textbf {\bibinfo {volume} {76}},\ \bibinfo {pages} {1089} (\bibinfo
  {year} {2000})}\BibitemShut {NoStop}%
\bibitem [{\citenamefont {Langrock}\ \emph {et~al.}(2005)\citenamefont
  {Langrock}, \citenamefont {Diamanti}, \citenamefont {Roussev}, \citenamefont
  {Yamamoto}, \citenamefont {Fejer},\ and\ \citenamefont
  {Takesue}}]{ref:Langrock_Fejer}%
  \BibitemOpen
  \bibfield  {author} {\bibinfo {author} {\bibfnamefont {C.}~\bibnamefont
  {Langrock \emph {et~al.}}},\ }\href@noop {} {\bibfield
  {journal} {\bibinfo  {journal} {Opt. Lett.}\ }\textbf {\bibinfo {volume}
  {30}},\ \bibinfo {pages} {1725} (\bibinfo {year} {2005})}\BibitemShut
  {NoStop}%
\bibitem [{\citenamefont {Vandevender}\ and\ \citenamefont
  {Kwiat}(2004)}]{ref:Vandevender_Kwiat_JMO}%
  \BibitemOpen
  \bibfield  {author} {\bibinfo {author} {\bibfnamefont {A.}~\bibnamefont
  {Vandevender}}\ and\ \bibinfo {author} {\bibfnamefont {P.}~\bibnamefont
  {Kwiat}},\ }\href@noop {} {\bibfield  {journal} {\bibinfo  {journal} {J. Mod.
  Opt.}\ }\textbf {\bibinfo {volume} {51}},\ \bibinfo {pages} {1433} (\bibinfo
  {year} {2004})}\BibitemShut {NoStop}%
\bibitem [{\citenamefont {Albota}\ and\ \citenamefont
  {Wong}(2004)}]{ref:Albota_Wong_upconversion}%
  \BibitemOpen
  \bibfield  {author} {\bibinfo {author} {\bibfnamefont {M.}~\bibnamefont
  {Albota}}\ and\ \bibinfo {author} {\bibfnamefont {F.}~\bibnamefont {Wong}},\
  }\href@noop {} {\bibfield  {journal} {\bibinfo  {journal} {Opt. Lett.}\
  }\textbf {\bibinfo {volume} {29}},\ \bibinfo {pages} {1449} (\bibinfo {year}
  {2004})}\BibitemShut {NoStop}%
\bibitem [{\citenamefont {Xu}\ \emph {et~al.}(2007)\citenamefont {Xu},
  \citenamefont {Ma}, \citenamefont {Mink}, \citenamefont {Hershman},\ and\
  \citenamefont {Tang}}]{ref:Xu_Tang}%
  \BibitemOpen
  \bibfield  {author} {\bibinfo {author} {\bibfnamefont {H.}~\bibnamefont
  {Xu \emph {et~al.}}},\ }\href@noop {} {\bibfield
  {journal} {\bibinfo  {journal} {Opt. Express}\ }\textbf {\bibinfo {volume}
  {15}},\ \bibinfo {pages} {7247} (\bibinfo {year} {2007})}\BibitemShut
  {NoStop}%
\bibitem [{\citenamefont {{Shah}}(1988)}]{ref:Shah_JQE}%
  \BibitemOpen
  \bibfield  {author} {\bibinfo {author} {\bibfnamefont {J.}~\bibnamefont
  {{Shah}}},\ }\href {\doibase 10.1109/3.124} {\bibfield  {journal} {\bibinfo
  {journal} {IEEE J. Quan. Elec.}\ }\textbf {\bibinfo {volume} {24}},\ \bibinfo
  {pages} {276} (\bibinfo {year} {1988})}\BibitemShut {NoStop}%
\bibitem [{\citenamefont {{Ma}}\ \emph {et~al.}(2011)\citenamefont {{Ma}},
  \citenamefont {{Bienfang}}, \citenamefont {{Slattery}},\ and\ \citenamefont
  {{Tang}}}]{ref:Ma_OE_11}%
  \BibitemOpen
  \bibfield  {author} {\bibinfo {author} {\bibfnamefont {L.}~\bibnamefont
  {{Ma}}}, \bibinfo {author} {\bibfnamefont {J.~C.}\ \bibnamefont
  {{Bienfang}}}, \bibinfo {author} {\bibfnamefont {O.}~\bibnamefont
  {{Slattery}}}, \ and\ \bibinfo {author} {\bibfnamefont {X.}~\bibnamefont
  {{Tang}}},\ }\href {\doibase 10.1364/OE.19.005470} {\bibfield  {journal}
  {\bibinfo  {journal} {Opt. Express}\ }\textbf {\bibinfo {volume} {19}},\
  \bibinfo {pages} {5470} (\bibinfo {year} {2011})}\BibitemShut {NoStop}%
\bibitem [{\citenamefont {{Kuzucu}}\ \emph {et~al.}(2008)\citenamefont
  {{Kuzucu}}, \citenamefont {{Wong}}, \citenamefont {{Kurimura}},\ and\
  \citenamefont {{Tovstonog}}}]{ref:kuzucu_wong}%
  \BibitemOpen
  \bibfield  {author} {\bibinfo {author} {\bibfnamefont {O.}~\bibnamefont
  {{Kuzucu \emph {et~al.}}}},\
  }\href {\doibase 10.1364/OL.33.002257} {\bibfield  {journal} {\bibinfo
  {journal} {Opt. Lett.}\ }\textbf {\bibinfo {volume} {33}},\ \bibinfo {pages}
  {2257} (\bibinfo {year} {2008})}\BibitemShut {NoStop}%
\bibitem [{\citenamefont {{Suchowski}}\ \emph {et~al.}(2010)\citenamefont
  {{Suchowski}}, \citenamefont {{Bruner}}, \citenamefont {{Arie}},\ and\
  \citenamefont {{Silberberg}}}]{ref:Suchowski_OPN_10}%
  \BibitemOpen
  \bibfield  {author} {\bibinfo {author} {\bibfnamefont {H.}~\bibnamefont
  {{Suchowski \emph {et~al.}}}},\
  }\href {\doibase 10.1364/OPN.21.10.000036} {\bibfield  {journal} {\bibinfo
  {journal} {Opt. Photon. News}\ }\textbf {\bibinfo {volume} {21}},\ \bibinfo
  {pages} {36} (\bibinfo {year} {2010})}\BibitemShut {NoStop}%
\bibitem [{\citenamefont {Flissikowski}\ \emph {et~al.}(2001)\citenamefont
  {Flissikowski}, \citenamefont {Hundt}, \citenamefont {Lowisch}, \citenamefont
  {Rabe},\ and\ \citenamefont {Henneberger}}]{ref:Flissikowski}%
  \BibitemOpen
  \bibfield  {author} {\bibinfo {author} {\bibfnamefont {T.}~\bibnamefont
  {Flissikowski \emph {et~al.}}},\ }\href {\doibase
  10.1103/PhysRevLett.86.3172} {\bibfield  {journal} {\bibinfo  {journal}
  {Phys. Rev. Lett.}\ }\textbf {\bibinfo {volume} {86}},\ \bibinfo {pages}
  {3172} (\bibinfo {year} {2001})}\BibitemShut {NoStop}%
\bibitem [{\citenamefont {Srinivasan}\ \emph {et~al.}(2008)\citenamefont
  {Srinivasan}, \citenamefont {Michael}, \citenamefont {Perahia},\ and\
  \citenamefont {Painter}}]{ref:Srinivasan17}%
  \BibitemOpen
  \bibfield  {author} {\bibinfo {author} {\bibfnamefont {K.}~\bibnamefont
  {Srinivasan \emph {et~al.}}},\
  }\href@noop {} {\bibfield  {journal} {\bibinfo  {journal} {Phys. Rev. A}\
  }\textbf {\bibinfo {volume} {78}},\ \bibinfo {pages} {033839} (\bibinfo
  {year} {2008})}\BibitemShut {NoStop}%
\bibitem [{\citenamefont {{Patel}}\ \emph {et~al.}(2010)\citenamefont
  {{Patel}}, \citenamefont {{Bennett}}, \citenamefont {{Farrer}}, \citenamefont
  {{Nicoll}}, \citenamefont {{Ritchie}},\ and\ \citenamefont
  {{Shields}}}]{ref:patel_Nphot2010}%
  \BibitemOpen
  \bibfield  {author} {\bibinfo {author} {\bibfnamefont {R.~B.}\ \bibnamefont
  {{Patel \emph {et~al.}}}},\ }\href {\doibase
  10.1038/nphoton.2010.161} {\bibfield  {journal} {\bibinfo  {journal} {Nature
  Photonics}\ }\textbf {\bibinfo {volume} {4}},\ \bibinfo {pages} {632}
  (\bibinfo {year} {2010})}\BibitemShut {NoStop}%
\bibitem [{\citenamefont {Flagg}\ \emph {et~al.}(2010)\citenamefont {Flagg},
  \citenamefont {Muller}, \citenamefont {Polyakov}, \citenamefont {Ling},
  \citenamefont {Migdall},\ and\ \citenamefont {Solomon}}]{ref:Flagg_PRL10}%
  \BibitemOpen
  \bibfield  {author} {\bibinfo {author} {\bibfnamefont {E.~B.}\ \bibnamefont
  {Flagg \emph {et~al.}}},\ }\href {\doibase
  10.1103/PhysRevLett.104.137401} {\bibfield  {journal} {\bibinfo  {journal}
  {Phys. Rev. Lett.}\ }\textbf {\bibinfo {volume} {104}},\ \bibinfo {pages}
  {137401} (\bibinfo {year} {2010})}\BibitemShut {NoStop}%
\bibitem [{\citenamefont {Pe'er}\ \emph {et~al.}(2005)\citenamefont {Pe'er},
  \citenamefont {Dayan}, \citenamefont {Friesem},\ and\ \citenamefont
  {Silberberg}}]{ref:Silberberg_PRL}%
  \BibitemOpen
  \bibfield  {author} {\bibinfo {author} {\bibfnamefont {A.}~\bibnamefont
  {Pe'er \emph {et~al.}}},\
  }\href@noop {} {\bibfield  {journal} {\bibinfo  {journal} {Phys. Rev. Lett.}\
  }\textbf {\bibinfo {volume} {94}},\ \bibinfo {pages} {073601} (\bibinfo
  {year} {2005})}\BibitemShut {NoStop}%
\end{thebibliography}
\end{document}